# Flexible MgO barrier magnetic tunnel junctions

Li Ming Loong[1], Wonho Lee[2], Xuepeng Qiu[1], Ping Yang[3], Hiroyo Kawai[4], Mark Saeys[5], Jong-Hyun Ahn[2]*, and Hyunsoo Yang[1]*

[1]*Department of Electrical and Computer Engineering, NUSNNI, National University of Singapore, 117576 Singapore*
[2]*School of Electrical and Electronic Engineering, Yonsei University, Seoul, 120-749, Korea*
[3]*Singapore Synchrotron Light Source, National University of Singapore, 5 Research Link, 117603 Singapore*
[4]*Institute of Materials Research and Engineering, 3 Research Link, Singapore 117602, Singapore*
[5]*Laboratory for Chemical Technology, Ghent University, Technologiepark 914, 9052 Ghent, Belgium*

* E-mail: ahnj@yonsei.ac.kr, eleyang@nus.edu.sg
[+] L.L. and W.L. contributed equally to this work.

## Abstract

Flexible electronic devices require the integration of multiple crucial components on soft substrates to achieve their functions. In particular, memory devices are the fundamental component for data storage and processing in flexible electronics. Here, we present flexible MgO barrier magnetic tunnel junction (MTJ) devices fabricated using a transfer printing process, which exhibit reliable and stable operation under substantial deformation of the device substrates. In addition, the flexible MTJ devices yield significantly enhanced tunneling magnetoresistance (TMR) of ~300 % and improved abruptness of switching, as residual strain in the MTJ structure induced by the fabrication process is released during the transfer process. This approach could be useful for a wide range of flexible electronic systems that require high performance memory components.

Flexible electronics has become the subject of active research in recent times, with studies exploring the fabrication of flexible transistors,[1] capacitors,[2] implantable medical devices,[3] and even magnetoresistive sensors.[4-6] In particular, memory devices are the fundamental component for data storage and processing for wearable electronics and biomedical devices,[3, 7-11] which require various functions such as wireless communication, information storage and code processing. Although a substantial amount of research has been carried out on organic resistive memory,[12, 13] as well as carbon material,[14-16] and inorganic thin film based memory,[17] there are still significant challenges in fabricating devices on soft substrates without sacrificing performance. Magnetoresistive random access memory (MRAM) based on magnetic tunnel junctions (MTJs) has been considered a promising storage element due to its exceptional merits, such as low power consumption, high speed, and unlimited read/write endurance.[18] While flexible alumina tunnel barrier MTJs have been reported,[4, 6] MgO based MTJ devices have been fabricated mostly on rigid and flat substrates,[19-21] and there have been no reports of flexible MTJs with MgO tunnel barriers yet. Nonetheless, as MgO barrier MTJs can yield significantly higher tunneling magnetoresistance (TMR) than their alumina counterparts, they are promising candidates for flexible electronics applications. In addition, it is more challenging to fabricate flexible MgO-barrier MTJs than their alumina counterparts, as the specific crystal structure of the MgO-barrier MTJs should not be compromised by the fabrication process. Hence, the successful fabrication of flexible MgO-barrier MTJs would be considered a milestone.

The emerging field dubbed "straintronics"[22-27] involves the integration of strain with spintronic devices including MTJs, where strain could be used to desirably manipulate spintronics phenomena in the devices. For example, the usage of strain generated by a ferroelectric or piezoelectric material to rotate the magnetization of an adjacent ferromagnetic layer via inverse magnetostriction (also known as the Villari effect) has been proposed as a novel magnetization switching method for applications such as MRAM.[28, 29] Hence, other than flexible electronics, another potential future direction for MTJ devices and applications could involve the incorporation of straintronics.

In this work, we present flexible MgO barrier MTJ devices fabricated using a transfer printing process, which exhibit reliable and stable operation under substantial deformation of the device substrates. We grow CoFeB/MgO/CoFeB MTJs on conventional, thermally-oxidized silicon substrates, release the MTJs from the substrates by etching away the underlying silicon, and then transfer and adhere the MTJs onto flexible polyethylene terephthalate (PET) substrates. Our flexible MgO barrier MTJs demonstrate improved performance on soft substrates by controlling the effect of strain on the devices. We report an approach to effectively improve device performance through the careful introduction of mechanical deformation in MTJs. The flexible MTJ devices yield significantly enhanced tunneling magnetoresistance (TMR) of ~300% and improved abruptness of switching, as residual strain in the MTJ structure induced by the fabrication process is released during the transfer process. In addition, the response and robustness of the flexible MTJs under strain are characterized in this work. The experimental work is complemented with quantum tunneling simulations. The results could provide useful insights for the design and engineering of novel MgO barrier MTJ-based straintronics as well as flexible electronics applications.

**Figure 1**a shows the $Co_{40}Fe_{40}B_{20}$/MgO/$Co_{40}Fe_{40}B_{20}$ MTJ film structure, which was deposited on Si/thermal $SiO_2$ (300 nm) substrates by magnetron sputtering at room temperature. Using photolithography and Ar ion milling, the film stack was patterned to form isolated MTJs with sizes ranging from 80 – 900 $\mu m^2$. Where necessary, the MTJ devices were post-annealed in an in-plane magnetic field of 0.055 T under ultra-high vacuum conditions. Using the four probe measurement technique, TMR measurements were performed on the MTJs at room temperature, before the substrate transfer process. Suspended MTJ devices were formed by lateral etching of the sacrificial Si layer using dry etching methods.[30] Then, the suspended stacks were transferred onto a PET substrate by dry transfer methods.[31] In addition, other approaches such as water-assisted transfer,[32] may be employed for transferring the devices if an assistive metal layer to create the desired crystal structure of the MgO-barrier MTJs can be formed.

It is well-known[33, 34] that thermal $SiO_2$ has intrinsic compressive stress of approximately −330 MPa. As shown schematically in Figure 1b, when the underlying Si substrate is etched away during the substrate transfer process, segments of the $SiO_2$ etch stop layer are released and relaxation of the intrinsic compressive stress occurs in these segments, thus imparting in-plane tensile strain to the overlying MTJ film stack. As the MTJs are never removed from the $SiO_2$ layer, they still retain this in-plane tensile strain even after the substrate transfer process. Figure 1c schematically depicts the changes in the atomic lattices of the different layers as a result of the strain induced by the substrate transfer process.

As shown in **Figure 2**a, where the same MTJ was measured before and after its transfer onto a flexible PET substrate, electrical measurements of the MTJs after the substrate transfer process yield a general increase in TMR, coercivities, and TMR loop squareness. Furthermore, Figure 2b, which summarizes the mean pre- and post-transfer TMR values as a function of pre-transfer annealing temperatures, shows an increase of the TMR for MTJs on PET that had been annealed at > 300 ºC. Figures 2a,b show that post-transfer TMR values can be enhanced to more than 200%. The results in Figure 2a,b could be attributed to the correlation between strain-enhanced TMR and coherent tunneling, where in-plane biaxial tensile strain has been found to increase the TMR of MTJs that exhibit coherent tunneling, as discussed later. In addition, the crystallization of the CoFeB ferromagnetic layers, which is required for coherent tunneling in the CoFeB/MgO/CoFeB structure, has been found to be complete only above 325 ºC.[35, 36] Hence, this is consistent with the data in Figure 2b, where the TMR seems to be enhanced post-transfer only for devices annealed at temperatures > 300 ºC.

The in-plane tensile strain imparted to the MTJ stack by the substrate transfer process was estimated using x-ray diffraction (XRD), as shown in Figure 2c. From the shift of the Cu (111) peak, the change in the out-of-plane lattice constant due to the transfer process was estimated, yielding out-of-plane strain of −0.22%. Using the Poisson's ratio of 0.34 for Cu,[37] as well as the equation for plane stress,[38] $\varepsilon_{zz} = -2\nu\varepsilon_{xx}/(1-\nu) = -2\nu\varepsilon_{yy}/(1-\nu)$, the in-plane tensile strain value was then estimated to be +0.21%. Moreover, as shown in Figure 2d, finite element analysis (FEA) was also used to estimate the strain imparted to the MTJs as a result of the release of the devices from the original Si substrate (see Supporting Information). In the simulation, the thermal $SiO_2$/MTJ device structure was modeled, and the relaxation of the intrinsic compressive stress in the thermal $SiO_2$ was simulated by applying an outward-directed pressure of 330 MPa on four side faces of the thermal $SiO_2$ layer, defined by their normals in Figure 2d as $\pm x$ and $\pm y$. Consequently, the MTJ device was stretched by the underlying thermal $SiO_2$ layer, giving rise to in-plane tensile strain of +0.2% in the Cu contact pads. Hence, the simulation results are fairly consistent with the XRD results obtained from Figure 2c.

The effects of the residual stress, released by the transfer process, on the devices can be clearly explained by comparison with those of in-plane biaxial tensile strain applied directly to the devices on the original Si substrate. For comparison, devices on a rigid Si substrate were subjected to measurements using a setup reported elsewhere,[39] where in-plane biaxial tensile strain was imparted to the MTJ devices at values of strain similar to those imparted by the transfer process. As shown in **Figure 3**a,b, the TMR, coercivities, and TMR loop squareness increased as the magnitude of the strain was increased, and at 0.15%, exhibited similar values and shape as those of devices transferred onto PET. Figure 3b summarizes the changes in TMR and coercivities ($H_C$) from Figure 3a as a function of the applied strain, where the increasing TMR has been attributed to the effects of strain on coherent tunnelling,[39] and the increasing coercivities of both the soft and hard magnetic layers can be attributed to the Villari effect. The results show that intrinsic stress relaxation by the transfer process (Figure 2a) has a similar enhancement effect on the TMR of the MTJs as directly applied external strain (Figure 3a).

Further experiments were performed to evaluate the response and durability of the post-transfer MTJ devices under substrate bending. The experimental setup used to apply uniaxial tensile and compressive strain to the flexible MTJ devices is shown in the inset of Figure 3f. For example, Figure 3c shows the effects of different levels of tensile strain on a MTJ transferred onto a PET substrate, where the uniaxial tensile strain was applied parallel to the easy axis of the MTJ. The TMR remains virtually constant, while the coercivities of the ferromagnetic layers increase as the magnitude of the uniaxial tensile strain increases, due to the Villari effect, indicating the tunability of the coercivities. Therefore, Figure 3c provides a gauge of the robustness of the devices under uniaxial tensile strain, as it retains its original properties when the strain is released (orange curve).

In addition, Figure 3d shows the effects of compressive strain on a post-transfer MTJ, where the uniaxial compressive strain was applied parallel to the initial easy axis of the MTJ. A significant reduction in TMR of ~ 25% was observed under a uniaxial compressive strain of −0.3%. This reduction can also be attributed to the Villari effect, since rotating the device in-plane 10° away from the initial easy axis almost completely restores the TMR to its original value of ~300%. Hence, the decrease in TMR due to uniaxial compressive strain was actually reversible, suggesting that the TMR is robust under uniaxial compressive strain. Due to the positive coefficient of magnetostriction of CoFeB, the application of uniaxial compressive strain along the initial easy axis of the MTJ effectively rotates the magnetic easy axis away from its original orientation, resulting in imperfect antiparallel alignment between the ferromagnetic layers of the MTJ, and the temporary decrease in TMR. Similarly, in the uniaxial tensile case (Figure 3c), the tensile strain reinforces the initial easy axis of the MTJ, and the TMR saturates though the magnitude of the strain increases because the magnetization is already saturated along this easy axis. Figure 3d illustrates the potential strain gauge application of the post-transfer devices,[40] as well as demonstrates the durability of the post-transfer devices under uniaxial compressive strain. Figure 3e summarizes the changes in TMR and coercivities under different levels of uniaxial tensile and compressive strain.

Measurements were also conducted to evaluate the durability of the devices under repeated substrate bending, and as a function of time. Figure 3f (circular symbols) provides a gauge of the robustness of a post-transfer MTJ that was repeatedly subjected to alternately uniaxial tensile and compressive flexes. As the devices are sometimes kinked after being transferred onto the second substrate (as shown in Figure 1a, some waviness is introduced to the devices when they are released from the original substrate), the initial increase in TMR shown in Figure 3f could be attributed to the "unkinking" of the transferred $SiO_2$/MTJ device segment as a result of the flexing. As random kinks may introduce undesirable strain to a device, the removal, or "straightening out", of such kinks could improve device performance. Continuing to repeatedly flex the device up to 40 times did not significantly alter its TMR, reflecting the durability of post-transfer devices under repeated strain. In addition, the device durability could potentially be enhanced by developing and incorporating a suitable corrugated flexible substrate structure to accommodate the strain.[5] Furthermore, Figure 3f (diamond symbols) shows the TMR of another post-transfer device, which was strained and tested repeatedly over a duration spanning more than two weeks. The TMR remained high and constant, further providing a gauge of the robustness of the post-transfer devices. The magnitude of applied strain was estimated using $\varepsilon \approx t_{substrate}/2R_C$,[41] where the substrate thickness ($t_{substrate}$) was 188 μm, and the radius of curvature ($R_C$) was obtained from optical images of the setup during measurements. The strain magnitude was also verified using FEA (Supporting Information). In addition, it is necessary to consider the strain gradient created by bending, which can induce flexoelectricity and flexomagneticity affecting the properties of device.[42, 43] The flexoelectric effect in the device could possibly affect the TMR values, and be an interesting area for future study. In contrast, the flexomagnetic effect, in which a material should be non-magnetic in its ground state and yet have strong magneto-elastic coupling, is insignificant becasue the materials used in our devices do not fulfil these criteria.

The transfer process is versatile, and can be used not only to fabricate flexible MgO barrier MTJs, but also to integrate MgO barrier MTJs with various substrates, thus removing the design constraint of having to grow the thin film structure on only certain types of substrates in order to safeguard the thin film quality. For example, besides PET, we demonstrate the transfer of MTJs onto various other substrates, such as glass, Al foil, polydimethylsiloxane (PDMS), and nitrile glove, as shown in **Figure 4**a-d. This versatility could facilitate the realization of various novel applications, such as wearable flexible sensors[11] and transparent electronics.[7] A typical TMR loop of a device post-transfer onto Al foil is shown in Figure 4e, demonstrating that the MTJs can still exhibit good performance even after being transferred onto a substrate other than PET. Moreover, Figure 4f compares the normalized mean TMR values pre- and post-transfer onto various substrates, where the post-transfer TMR is 1.38 times higher than the pre-transfer TMR, on average. The average enhancement factor is comparable to that reported for the case where similar strain was applied directly to the devices,[39] suggesting that the enhancement can be attributed to the effects of strain on the quantum tunneling.

In order to gain insight into the effect of biaxial strain due to the substrate transfer process on the conductance and the TMR, non-equilibrium Green's function (NEGF) quantum transport calculations were performed. The $k_∥$-resolved transmission spectra, $T(E_F)$ for the parallel (P) and antiparallel (AP) configuration were calculated for biaxial *xy*-strain ranging from -1% to 1% (**Figure 5**a), where negative and positive strain correspond to compressive and tensile strain, respectively. The change in the conductance for the P and AP configuration and the resulting change in the TMR are shown in Figure 5c. For the P configuration, conduction is dominated by states at the $\Gamma$ point, in agreement with previous studies.[39, 44-47] The $T(E_F)$ at the $\Gamma$ point decreases with increasing tensile strain, and hence the conductance decreases monotonously with strain, as shown in Figure 5c (black squares). For the AP configuration, however, several channels away from the $\Gamma$ point contribute to the tunneling transport,[39, 44-47] and the effect of strain on the conductance depends on the location in the Brillouin zone. The transmission coefficient $T(E_F)$ for states close to the $\Gamma$ point decreases significantly with tensile strain, similar to but faster than the $T(E_F)$ for the P configuration, but for tensile strains close to 1%, the $T(E_F)$ for states somewhat further away from the $\Gamma$ point shows an opposite trend and gradually increases. The overall AP conductance still decreases with strain, but the competition between the different channels causes the decrease to saturate and possibly reverse for larger tensile strains (Figure 5c, red triangles). Since the change in the AP conductance is initially much larger than the change in the P conductance, the TMR increases with strain, but saturates for tensile strains close to 1% (Figure 5c, blue circles). This result qualitatively agrees with the trend observed in the experiments (Figure 2a and 3a).

The change in the overall conductance and in the TMR of the junction caused by biaxial strain hence results from a competition between different effects, making a quantitative description of the effect of strain on the TMR challenging. Our simulations clearly illustrate this subtle balance for the AP configuration, and hence indicate a possibility of strain engineering to optimize the TMR.

We demonstrate functional, flexible MgO barrier MTJs for the first time using a substrate transfer process. Furthermore, we observe an improvement in the MTJ properties after the transfer process, which could be attributed to intrinsic stress relaxation and strain-enhanced coherent TMR. Our results provide a proof of concept for flexible MgO barrier MTJs, which are promising for various novel applications, including sensors and data storage devices.

**Experimental Section**

*Thin film and device fabrication:* The sputter-deposited film stack structure was Si substrate/SiO$_2$ (300)/Ta (5)/Ru (20)/Ta (5)/Co$_{40}$Fe$_{40}$B$_{20}$ (6)/MgO (2)/Co$_{40}$Fe$_{40}$B$_{20}$ (4)/Ta (5)/Ru (5) (thicknesses in nm). All the metal layers were deposited using dc sputtering, while the MgO tunnel barrier and SiO$_2$ encapsulation were deposited using rf sputtering. The sputtering pressures for the different layers were in the range of 1 − 3 mTorr. The first photolithography step was performed followed by Ar ion milling to define completely isolated mesas of the full film stack. Then, the photoresist was stripped off and the second photolithography step was performed to define the MTJs. The subsequent Ar ion milling process was monitored using secondary ion mass spectroscopy (SIMS), enabling the vertical milling to be stopped after the MgO barrier had been passed. Low-angle milling was then performed to remove undesirable sidewalls, where the ion beam was incident at 20° from the sample plane. Next, the devices were encapsulated with 50 nm of SiO$_2$, lift-off was performed, and the third photolithography step was carried out to define the contact pads. A Ta (4 nm)/Cu (100 nm) contact pad structure was sputter-deposited onto the samples, and lift-off was performed to complete the fabrication process.

*Transfer process:* The outside parts of MTJs ribbon (300 nm-thick SiO$_2$ layer) was vertically etched with a buffered oxide etch (BOE) to expose the underlying bulk Si. Next, lateral etching of the sacrificial Si layer was carried out by inductively-coupled-plasma reactive ion etching (ICP-RIE) with SF$_6$ and CHF$_3$ gases for 35 min to form the suspended structures. When a PDMS stamp (agent to base ratio ~ 1:10) is attached to the suspended MTJs and detached, the whole array of suspended MTJs was lifted off from the original substrate at once. Then, they are transferred to various target substrates coated with SU-8 (300 nm-thick, semi-cured by UV exposure of 200 mJ/cm$^2$ for 10 sec), namely glass, Al foil, PDMS, nitrile gloves, and PET substrates due to difference of adhesion forces. The stress/strain distribution in MTJs are not changed during the transfer step. After transfer printing the MTJ devices, annealing was performed at 65 °C for 30 minutes to fully cure the SU-8 adhesive layer. The transferred devices were stably fixed on the substrate after the full curing process.

*Quantum tunneling simulations:* The tunneling junction was modelled as a 10-layer 2 nm-thick MgO(001) barrier sandwiched between semi-infinite Fe(100) contacts (Figure 5b). The electronic structure of the junction was described by an extended Hückel molecular orbital (EHMO) Hamiltonian, as implemented in Green,[48, 49] and standard EHMO parameters were used.[39, 44, 50] Biaxial strain was introduced in the *x* and *y* direction and the change in the *z*-direction was obtained from the Poisson's ratios of 0.19 for MgO and 0.37 for Fe.[39, 51] An average Poisson's ratio of 0.28 was used for the Fe(100)-MgO(001) interface distance. The experimental Fe lattice constant of 2.86 Å and a Fe(100)-MgO(001) distance of 2.16 Å were used for the unstrained case.[45] The $k_{\parallel}$-resolved transmission spectra at the Fermi level, $T(E_F)$, were computed for a fine (100 × 100) grid covering the transverse Brillouin zone.

**Supporting Information**
Supporting Information is available online.

**Acknowledgements**
This research is supported by the National Research Foundation, Prime Minister's Office, Singapore under its Competitive Research Programme (CRP Award No. NRF-CRP12-2013-01). J.-H.A. acknowledges financial support by the National Creative Research Laboratory (2015R1A3A2066337) through the National Research Foundation of Korea (NRF) and the Yonsei University Future-leading Research Initiative. P.Y. is supported by SSLS via NUS Core Support C-380-003-003-001. H.K. acknowledges the A*STAR Computational Resource Centre (A*CRC) for the computational resources and support. H.Y. is a member of the Singapore Spintronics Consortium (SG-SPIN).

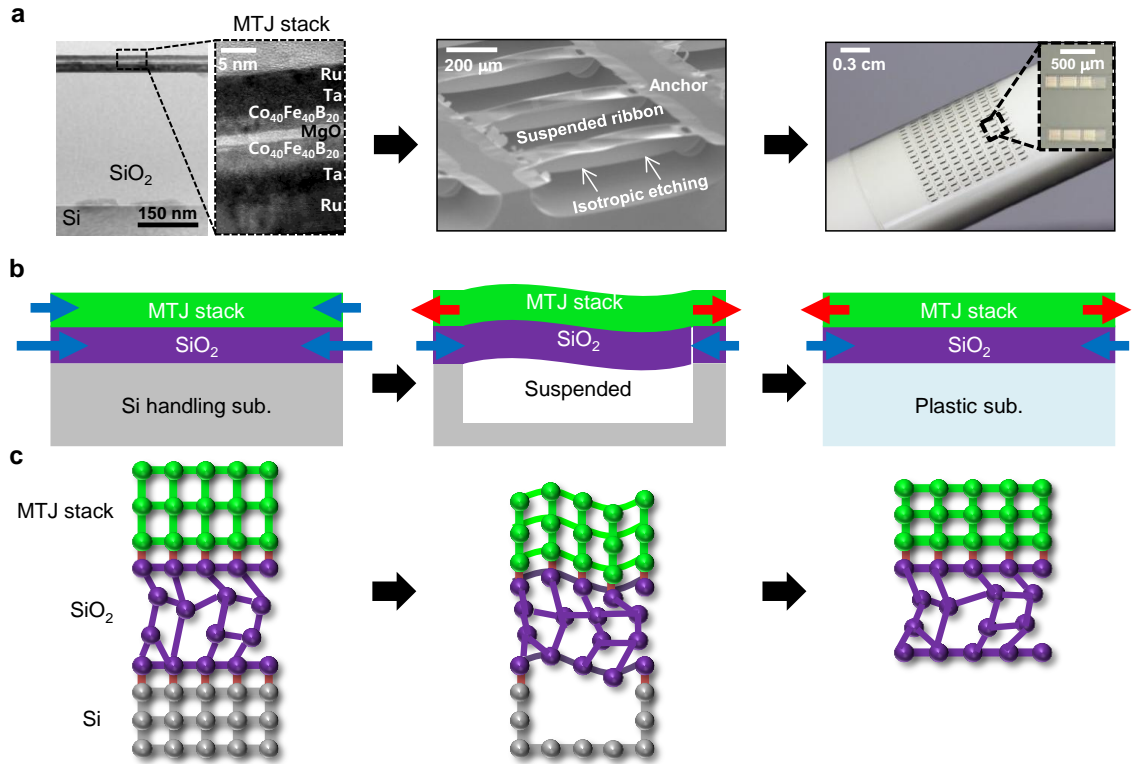

**Figure 1.** a) The crystallinity of the MgO tunnel barrier and the adjacent CoFeB ferromagnetic layers in the annealed MTJ film stack on the original Si/SiO$_2$ substrate was verified by transmission electron microscopy (TEM). The devices were subjected to Si undercut etching, as shown in the scanning electron microscope (SEM) image. The devices were then transferred onto a PET substrate, as shown in the optical images. b) Schematic diagram showing the transfer process, where the arrows represent the intrinsic stresses in the film layers. c) Schematic diagram showing the changes in the atomic lattices and strain of the different film layers.

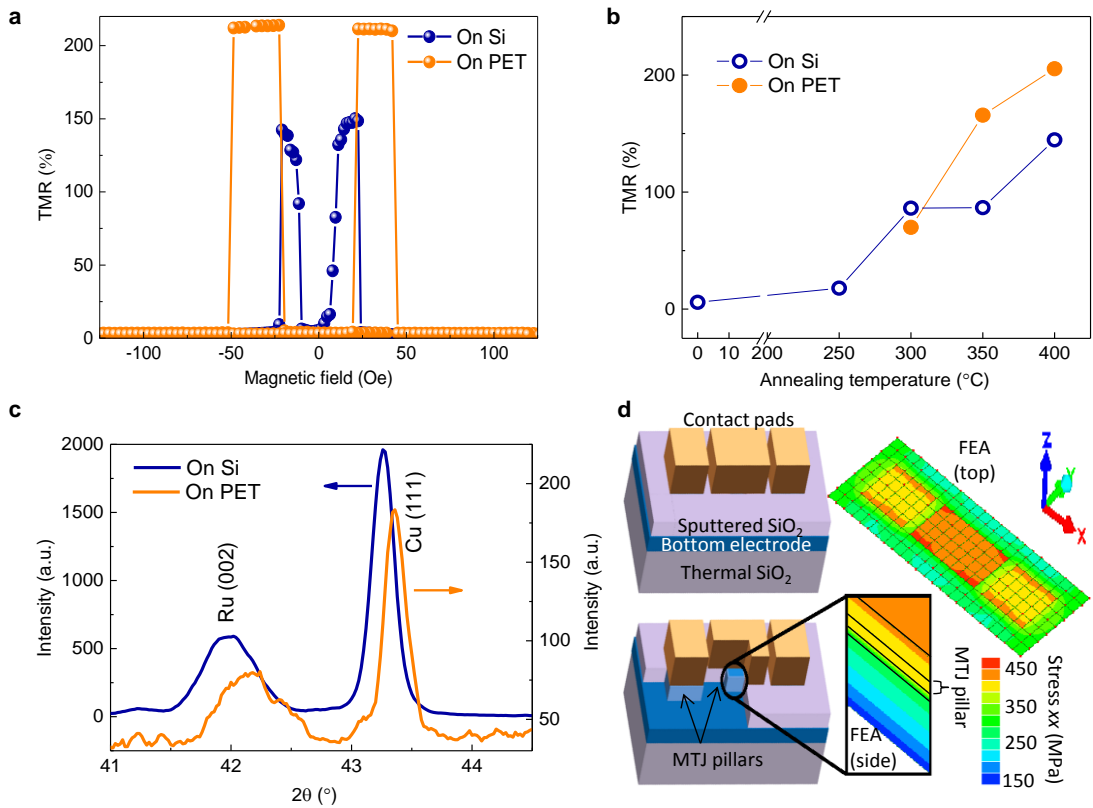

**Figure 2.** a) TMR loops of a device before and after the transfer onto PET, showing enhanced device performance after the transfer. b) The mean TMR of fabricated devices for different annealing temperatures. The corresponding mean TMR values for devices transferred onto PET are included for some of the annealing temperatures, for comparison. c) XRD data from devices before and after the transfer, suggesting in-plane biaxial tensile strain of 0.2 % due to the transfer. d) Schematic diagrams of the MTJ device structure, including a cutaway schematic showing the MTJ pillars, which would otherwise be obscured by the contact pads in the finished devices. The top and side views of the FEA simulation results for a MTJ device after its release from the original Si substrate are also shown, specifically the distribution of the *xx* component of the stress due to the release.

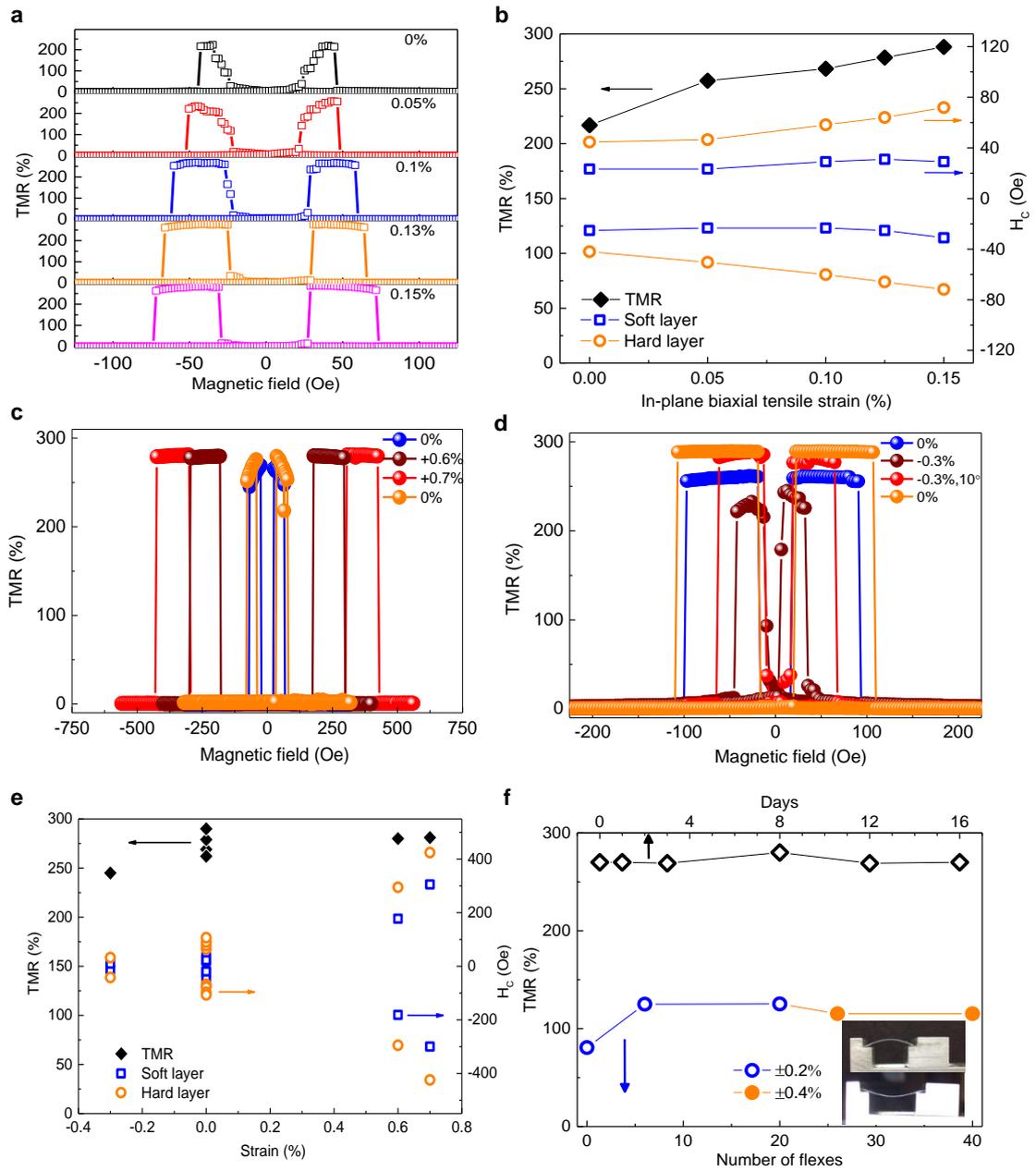

**Figure 3.** a) TMR loops for a device on a conventional Si/SiO$_2$ substrate, when subjected to increasing levels of in-plane biaxial tensile strain. b) Summary of the changes in TMR and the coercivities ($H_C$) of the magnetically soft (ferromagnetic layer with a lower $H_C$) and hard (ferromagnetic layer with a higher $H_C$) layers of the device in Figure 3a, as the in-plane biaxial tensile strain increases. TMR measurements of a post-transfer MTJ on PET being subjected to different levels of uniaxial (c) tensile and (d) compressive strain. e) Summary of the changes in TMR and the $H_C$ of the magnetically soft and hard layers for Figure 3c and 3d, as the in-plane uniaxial strain is changed. f) The data corresponding to the bottom *x*-axis are from a post-transfer MTJ on PET after 20 flexes at ±0.2% strain (alternately uniaxial tensile and compressive) followed by another 20 flexes at ±0.4% strain. The data corresponding to the top *x*-axis are from another device, which was re-tested several times over a given duration. The inset shows the experimental setup for applying uniaxial tensile and compressive strain.

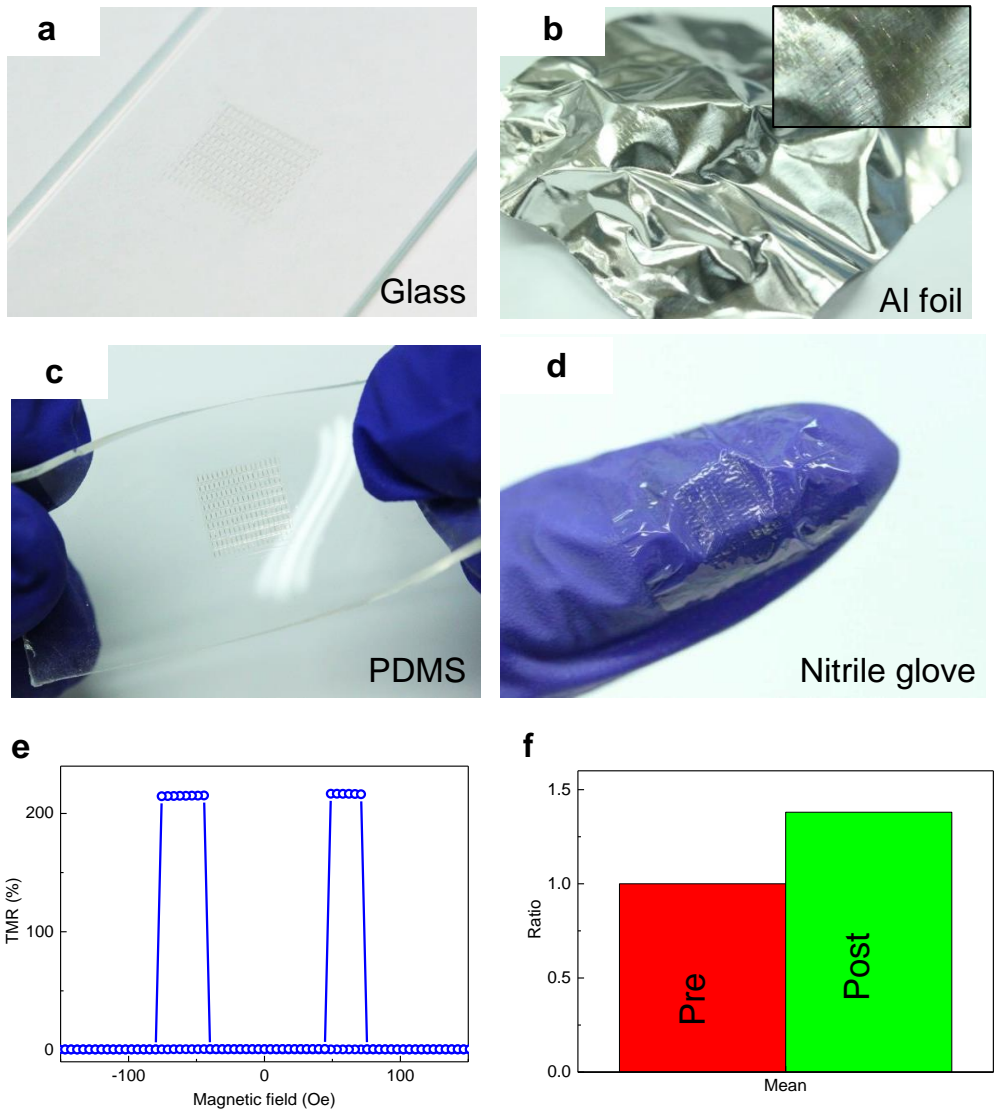

**Figure 4.** Optical images of MTJs transferred onto (a) glass, (b) Al foil, (c) PDMS, and (d) nitrile glove. The dimensions of each isolated MTJ mesa were 150 × 570 μm. e) TMR loop of a device post-transfer onto Al foil. f) Normalized mean TMR values taking into account data from different device batches corresponding to the various post-transfer substrates.

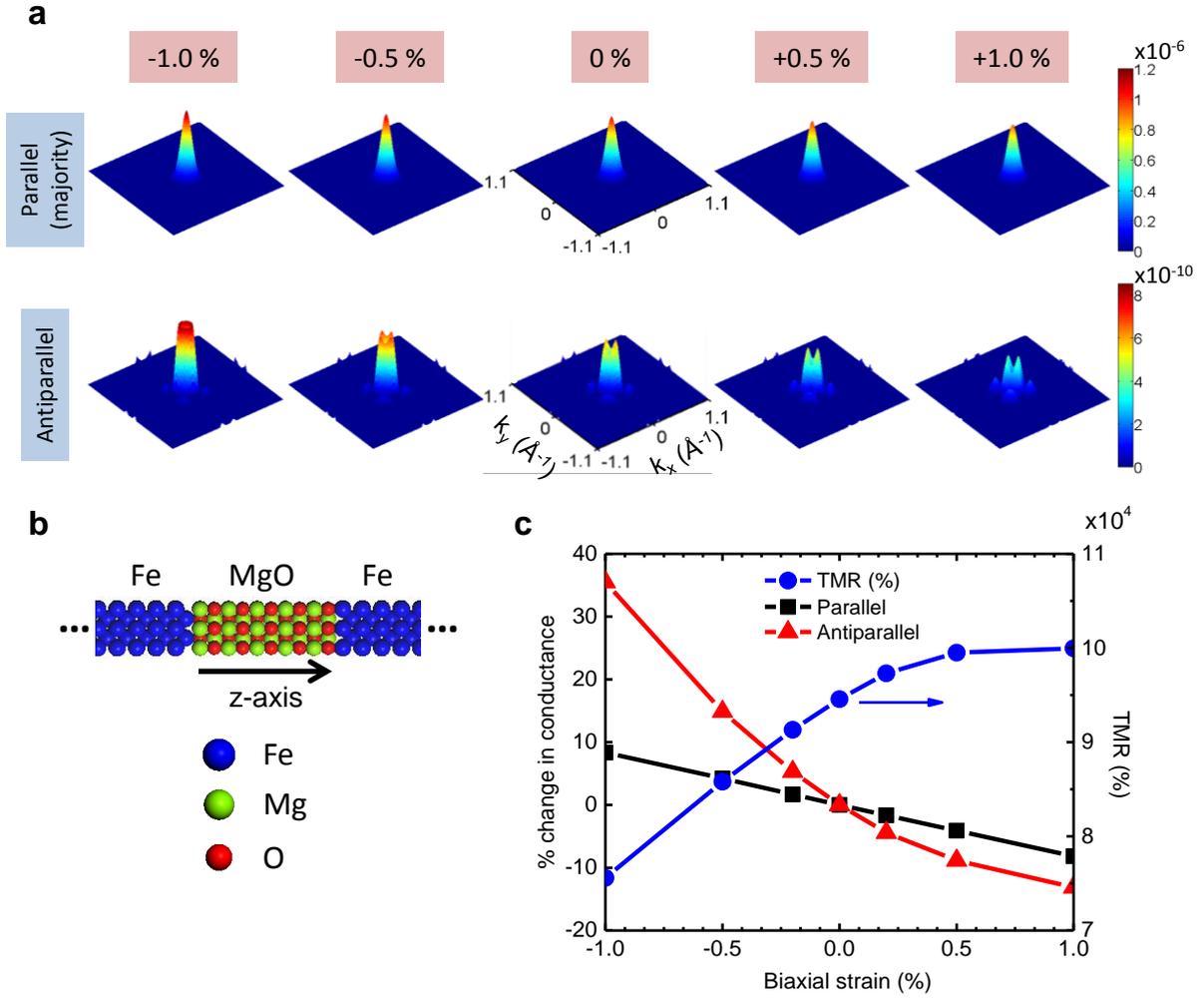

**Figure 5.** a) Calculated *k*-resolved transmission coefficients, $T(E_F)$, over the transverse Brillouin zone (-1.1 Å$^{-1}$ ≤ $k_x$ ≤ 1.1 Å$^{-1}$, -1.1 Å$^{-1}$ ≤ $k_y$ ≤ 1.1 Å$^{-1}$), for the majority states in the P and AP configuration for a range of biaxial *xy*-strain values. b) Fe/10-layer MgO/Fe model used to simulate the quantum transport, where the *z*-axis is perpendicular to the MgO layers. c) Percentage change in the conductance for the P and AP configuration and TMR ratio as a function of biaxial *xy*-strain. The Fermi level is located 4.3 eV above the MgO valence band edge.